# Maxwell's Demon Is Foiled by the Entropy Cost of Measurement, Not Erasure

R. E. Kastner

University of Maryland, College Park; rkastner@umd.edu

I dispute the conventional claim that the second law of thermodynamics is saved from a "Maxwell's Demon" by the entropy cost of information erasure, and show that instead it is measurement that incurs the entropy cost. Thus Brillouin, who identified measurement as savior of the second law, was essentially correct, and putative refutations of his view, such as Bennett's claim to measure without entropy cost, are seen to fail when the applicable physics is taken into account. I argue that the tradition of attributing the defeat of Maxwell's Demon to erasure rather than to measurement arose from unphysical classical idealizations that do not hold for real gas molecules, as well as a physically ungrounded recasting of physical thermodynamical processes into computational and information-theoretic conceptualizations. I argue that the fundamental principle that saves the second law is the quantum uncertainty principle applying to the need to localize physical states to precise values of observables in order to effect the desired disequilibria aimed at violating the second law. I obtain the specific entropy cost for localizing a molecule in the Szilard engine and show that it coincides with the quantity attributed to Landauer's principle. I also note that an experiment characterized as upholding an entropy cost of erasure in a "quantum Maxwell's Demon" actually demonstrates an entropy cost of measurement.

## 1. Introduction

This paper addresses the notorious problem of "Maxwell's Demon," a hypothetical microscopic being who is assumed able to make use of his knowledge of system microstates to effect a reduction in entropy of a system, thus violating the second law of thermodynamics. I do not presume to offer here a comprehensive background of that much-discussed issue. The current paper's focus is on the conventional claim, due to Bennett (1988), that the Demon is thwarted by the need to erase his memory. I critique that claim, and show that the entropy cost that thwarts the Demon comes instead from the need to measure specific observables to obtain the values needed by the Demon to accomplish his manipulations. Section 2 provides some background for the issue discussed here, as well as some definitions and conceptual clarifications that are helpful in situating the proposed solution.  Section 3 presents the  specific calculation that yields the entropy increase. Section 4 discusses aspects of the debate that have led to what is arguably a dysfunctional focus on computational notions and neglect of crucial quantum-level considerations concerning measurement. Section 5 summarizes and concludes.

## 2. Background

Recent tradition in addressing the problem of Maxwell's Demon has been to center the discussion around Landauer's Principle--the claim that in computational processes, information erasure is always accompanied by an entropy cost (Landauer, 1961). However, this paper  will not discuss Landauer's Principle (except to note the irrelevance of its computational form to the issue of Maxwell's Demon). This is because, as we shall see, there is no need to invoke computation, information storage, or erasure in order to address and fully resolve the challenge of the Demon.  Introducing these computational concepts have arguably served to deflect discussion to secondary issues (such as whether thermodynamic irreversibility corresponds to computational/logical irreversibility) and thus to divert attention from the essential issue: **can a microscopic being really manipulate gas molecules in order to achieve a disequilibrium state used to violate the second law**? It turns out that we don't actually need any computational concepts to address this question, and arguably such concepts have obscured the essential physics. All we need is the idea that one must obtain values of observables for physical systems in order to manipulate them to some desired end, which is what the Demon is doing. When we analyze that requirement for the quantum case, we find the heretofore missing cost of entropy arising from measurement.

In preparation for that result, derived in Section 3, it is  important to distinguish between  the usual classical (or quasi-classical) treatment and an accurate quantum-level treatment. Recall that in classical statistical mechanics, probability distributions characterizing macrostates arise from a model in which a system is assumed to possess determinate properties (such as occupying phase space points), but some designated observer is simply ignorant of those possessed properties. This is termed *epistemic uncertainty*: it quantifies *ignorance of some observer* about some actually-possessed, well-defined property or properties. Classical epistemic uncertainty is to be contrasted with intrinsic indeterminacy of a system's properties typical of quantum states, the latter being reflected for example in the variance of a Gaussian wave packet. Intrinsic indeterminacy can of course also be quantified by a probability distribution, e.g., $|\Psi(x)|^2$, but this distribution does not describe epistemic uncertainty in that it is not based on some observer's ignorance, but applies to the system itself. As such, it is often called *ontological* *uncertainty*.[1] The latter is subject to the constraints of the Heisenberg uncertainty principle,[2] with crucial physical ramifications for the entropy of the system, as will be demonstrated in what follows.

Specifically, the thesis of the present work is that, for real physical systems such as gas molecules-- which are quantum systems--we (or a Demon) are not just "ascertaining" the requisite states in the sense of remedying our ignorance about pre-existing values of observables, but instead are *creating* them (or at least bringing them about) via measurement, and that is associated with an entropy cost as quantified by the entropic form of the Heisenberg Uncertainy Relation, to be introduced below. Our approch thus refutes the fundamental assumption of the Bennettian tradition that the Demon's measurement process can be modeled as mere copying of some outcome-state that itself was arrived at without entropy cost.

The basic starting point is to note that the Heisenberg uncertainty principle has direct implications for the entropy of a system. Consider a system, such as a gas molecule, described by a position-wavefunction $\psi(x)$ and conjugate momentum-wavefunction $\varphi(p)$. The relation between these two wavefunctions in terms of "information" arises directly from the Heisenberg uncertainty relation, and takes the form

$$-\int_{-\infty}^{\infty}[\varphi(p)]^2 ln[\varphi(p)]^2 dp - \int_{-\infty}^{\infty}[\psi(x)]^2 ln[\psi(x)]^2 dx \;\geq\; ln\,\frac{e}{2} \quad (1)$$

This expression was pioneered by Weyl (1928) and discussed by Hirschman (1957) and Leipnik (1959):[3] In addition, a careful study of a similar "entropic uncertainty principle" and its applicability to the evolution of quantum systems has been undertaken by Geiger and Kedem (2022).

In view of the fundamental definition of thermodynamic entropy in terms of energies (rather than positions), the quantity properly indentified with thermodynamic entropy is the first term on the left hand side which depends on the momentum distribution, i.e.:

---

[1] Ontological uncertainty may be regarded as a form of "physical information", but is not necessarily equivalent to thermodynamic entropy. We discuss that issue in connection with eqn. (2) and further in Section 3. We agree that "information" is a generalization of thermodynamic entropy, as discussed for example in Nosonovsky (2010).

[2] While hidden variables quantum theories, such as the DeBroglie-Bohm pilot wave theories, view position uncertainty as epistemic, the strong coupling of those positions to the quantum state (via the "quantum potential") makes the uncertainty based on the state ontologically consequential. Thus, the presumed determinacy of Bohmian corpuscle positions is not representable by a classical phase space description.

[3] While the differential form is known to lead in principle to unphysical negative values, we can eliminate those by reference to a discretized version that takes into account that no real system is ever infinitesimally localized. (This is the dimensionless form, in which a term *ln h* is subtracted out.)

$$S = -k\int_{-\infty}^{\infty}[\varphi(p)]^2 ln[\varphi(p)]^2 dp \qquad (2)$$

We can also quantify S directly in terms of the variance of the distribution $|\varphi(p)|^2$ through the relation

$$S(\sigma) = \frac{k}{2} ln\,(2\pi\sigma^2 e) \qquad (3)$$

(cf. Leipnik 1959).

With these tools and concepts, we turn now to the entropy cost of measurement that thwarts the Demon from his goal of violating the second law.

3. Measurement as the source of entropy increase

We begin this section by observing that real gas molecules are quantum systems, and as such are subject to (1) Thus, any localizing measurement required to implement work extraction must take into account the uncertainty principle, and we show that this incurs an entropy cost of the same amount as that obtained in Landauer's Principle, but without any step involving erasure of stored computational data.

3.1 Entropy cost of measurement

Let us consider the Demon in the form of Szilard's engine, treated as a gas molecule in a chamber from which work can be extracted once it is localized to one side or the other and allowed to expand against a piston.

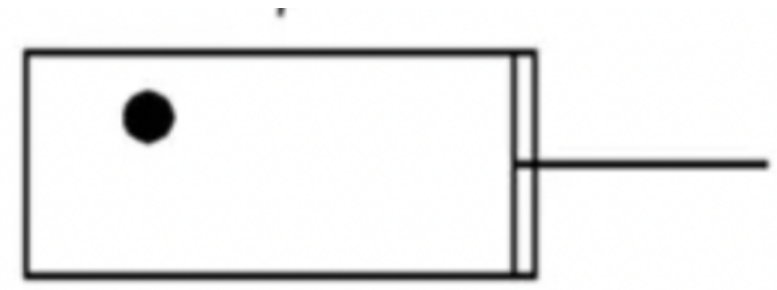

Figure 1. Typical depiction of gas molecule in Szilard's Engine.

First, let us note that the typical depiction of a gas molecule as a well-localized particle, shown in Figure 1, is misleading; this is an aspect of the persisent but inapplicable classical idealization that has long obscured the central role of measurement in the entropy cost of attempting to violate the second law. In fact, the molecule obeys the uncertainty principle and is not localized on one or the other side of the container as depicted. Thus, inserting the partition is actually the measurement that localizes it. That, in a nutshell, is the key point habitually overlooked in extant analyses. The molecule is more appropriately modeled as transitioning among energy eigenstates, as in the Boltzmann distribution. These are sinusoidal and for room temperature highly non-localized, as in Figure 2:

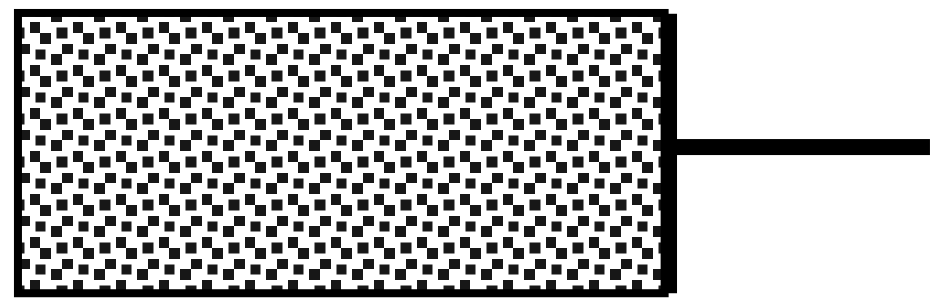

Figure 2. The gas molecule is not localized prior to partition insertion.

Allowing for decoherence effects, let us approximate the molecule's position wavefunction *ψ(x)* by a Gaussian. The Gaussian yields the smallest possible entropy increase as in (1), so an energy eigenstate would yield a larger increase. Thus this treatment leaves no 'loopholes' exploitable by considering different probability distributions, and is fully general in that sense.

The probability density $|\psi(x)|^2$ has a variance $\sigma_x^2$ initially of the order of the (squared) length L of the box: $\sigma_{x,i}^2 \sim L^2$. (We consider the 1-D situation for simplicity and without loss of generality.) Then the initial variance of the conjugate momentum, applying to $\varphi(p)$, is $\sigma_{p,i}^2 \sim \frac{\hbar^2}{4L^2}$. After we place the partition in the center, the position uncertainty is halved and variance has been reduced to $\sigma_{x,f}^2 \sim \frac{L^2}{4}$ and thus the momentum variance has increased to $\sigma_{p,f}^2 \sim \frac{\hbar^2}{L^2}$. The difference in initial and final entropies of the molecule can be found using the definition (2) and the expression relating entropy to variance (3). $S(\sigma) = \frac{k}{2} ln\ (2\pi\sigma^2 e)$While (3) depends on the units of σ, we only require the difference of the initial and final entropies in which the scaling effect of units cancels out; thus, substituting the above initial and final values for $\sigma_p^2$ in (3), we find:

$$\Delta S = S_f - S_i = \frac{k}{2} ln\ \frac{\sigma_f^2}{\sigma_i^2} = \frac{k}{2} ln\ 4 = k\ ln\ 2 \qquad (4)$$

Thus, the entropy cost for cutting the volume in half, and thereby bringing the system to a lower entropy state, is precisely the same as that appearing in Landauer's Principle and exactly compensates the reduction in entropy due to the smaller volume occupied by the system. And it is here where we resolve the alleged paradox of Szilard's engine: *we pay an entropic price for the work we have extracted, by the need to localize the* *molecule on one side or the other. And we pay that price before we even find out which side the molecule is on.*

It may be noted that insofar as Landauer's Principle (codifying an entropy cost for information erasure) is valid, it arises from a form of "erasure" that is about the narrowing of a probability distribution over values of  observables conjugate to 4-momentum. Indeed, (1) and (2) can be seen as another way of deriving Landauer's principle, but here it is in connection with quantum measurement and the intrinsic spread ("information") of wavefunctions, as opposed to computational bits in the classical sense. So the irony perhaps is that one can see Landauer's principle as applicable, but not to erasure in the sense of resetting a memory, since one does not need to refer to any sort of computational process or memory storage capacity in order to obtain the entropy cost saving the second law. The physically admissible grounding of Landauer's principle is found only in the narrowing of ontological uncertainty in position-related quantities.[5]

The above result is perhaps another way to look at the entropy cost pointed to in Norton (2025) for bringing a molecular component of a larger system from a higher to a lower entropy state. Norton formulates this as follows (where $\Omega$ refers to the number of complexions for the larger system):

> "In its briefest form, the relation just is Boltzmann's celebrated relation "S = k log $\Omega$" between entropy S and probability $\Omega$. If a thermodynamic process carries a system from state "1" to state "2," the driving entropy increase between the two states is $\Delta S = S2 – S1$. It relates to the probability of successful completion, $\Omega_2$,  [in which the entropy of a component, such as a molecule, is reduced]  and the probability that a fluctuation reverts the system to its initial state, $\Omega_1$, by $\Delta S = k \log (\Omega_2/\Omega_1)$. Thus, a probability ratio favoring completion in molecular-scale processes can only be enhanced by a dissipative increase in entropy creation $\Delta S$. The outcome is that, independently of any entropy cost associated with erasure or the logic implemented, there is an inevitable entropy cost associated with the suppression of fluctuations." (Norton 2025, 6)

In the present context, a process (insertion of the partition) forcing the molecular system out of equilibrium is compensated by an entropy cost of an amount given by (4), which matches the result from Norton's identified quantity, $\Delta S = k \log (\Omega_2/\Omega_1)$.  In effect the partition is both the agent of the creation of a disequilibrium state and of the suppression of fluctuations, and its insertion is indeed dissipative. Our calculation thus coincides with Norton's formulation, although it arises from a somewhat different consideration. This perhaps suggests a deep correspondence between classical statistical mechanics--which gets off the ground from a fundamental probability postulate not found in deterministic evolution--and non-deterministic (non-unitary) measuring interactions at the quantum level. The latter serve to momentarily localize particles and thus increase their momentum spread (entropy). This ubiquitous localizing process can therefore be identified as the physical source and justification of the fundamental probability postulate of classical statistical mechanics. Indeed, Brownian motion, which is the observable hallmark of fluctuations, arises only from genuinely Markovian dynamics in a system, which is arguably not obtained from classical deterministic evolution. The empirical fact of Brownian motion thereby suggests that non-unitary interactions are physically real at the quantum level (Kastner 2017). In any case, the result (4) concerning the entropy cost of measurement stands, irrespective of any interpretive conclusions one wishes to draw from the demonstrated entropy cost of measurement.

---

[5] Experimental "confirmations" of Landauer's principle are supported by increases in momentum/energy uncertainty in processes that amount to measurement; i.e., narrowing of position-related uncertainties, not by computational considerations involving "erasure" of epistemic uncertainty.

To summarize: a key step in the operation of a system like Szilard's engine, involving a single gas molecule in a chamber, is the insertion of a partition that localizes it to one side or another. A standard assumption in the current debate is that there is no entropy cost associated with this step and that a possible entropy cost of measurement only arises for the step in which an experimenter must learn which side of the partition the gas molecule is on. The latter is the step for which Bennett's keel-key device is intended (Bennett, 1988; however ineffective it would be, as pointed out by Norton, 2017). However, that standard analysis takes place under classical idealizations, which do not apply to a real gas molecule. It is a quantum system that must be described by a wave function. This is the case even if we assume an effectively "classical" partition function for the molecule, such as the Boltzmann distribution over energies. For no matter the quantum state of the molecule, it must undergo an entropy increase upon localization to one side or the other, with lower bound given by (4).

### 3.2 A possible objection refuted

We now address a possible objection to the above treatment: namely, questioning (2) as the appropriate characterization of thermodynamic entropy. Such an objection would arise from two possible sources:

(i) the conventional counting of phase space microstates towards entropy;

(ii) the information-theoretic tradition of equating Shannon information (as a general quantity) to thermodynamic entropy.

Regarding (i): due to the uncertainty principle, real quantum systems do not occupy phase space points, so that phase space volumes are only a classical approximation. Because only differences in entropy are directly measurable, this overcounting has not raised problems for empirical correspondence; however, it is not physically valid for quantum systems and therefore cannot be used as a basis for objection to the definition (2). Regarding (ii): the essential point is that thermodynamic entropy is fundamentally defined in terms of energy-related quantities, *not position*. Clausius thermodynamic entropy is $S = Q/T$. Position appears nowhere here. The volume factor that enters into the partition function for an ideal gas in the thermodynamic limit arises from the dependence of the energy levels on the dimension of the box. The volume factor does not arise from any need to count phase space position microstates.

Furthermore, it will not do to assert that "position information" constitutes thermodynamic entropy S, based on an insufficiently critical tradition of identifying Shannon "information" entropy with the Clausius thermodynamic entropy. While the two may be related under specific conditions, as discussed for example in Kastner and Schlatter (2024), the fact that position appears nowhere in the original quantification of the Clausius entropy, along with the fact that momentum (energy) and position are incompatible observables, forces the conclusion that one cannot legitimately count not-actually occupied position microstates proportional to container volume V and add them to $\Omega$ along with energy microstates. The only microstates legitimately counted are those contributing to quantities of heat Q and temperature T, in accordance with the Clausius definition of thermodynamic entropy $S = Q/T$.

A further observation might help to clarify this issue: it is a common practice to calculate the entropy change for a free expansion of an ideal gas by way of a reversible process, so that for example an expansion doubling the size of the gas results in an entropy change $\Delta S = Nk \ln 2$. This is obtained from the reversible process by calculating the work done and equating that to the heat exchanged with the environment to get $\Delta Q = T\,\Delta S$. But in a free expansion, no work is done on the environment and no heat is exchanged with the environment, yet the definition of entropy S in terms of heat Q remains. Thus for the irreversible process as well, one can say that a quantity of heat $\Delta Q = Nk\,T \ln 2$ applies to the change in entropy. What is this quantity of heat physically? It quantifies the net energy (heat) flow from the smaller volume to the larger one. The point of noting this is that it clarifies that thermodynamic entropy S *must always be reducible to heat flow* to comply with its basic definition. Thus, quantifying entropy does not consist solely in counting general microstates, especially if such microstates do not correspond to some transferred quantity of energy.

## 4. Critique of the 'erasure' tradition

### 4.1 Unphysical classical idealizations obscure the essential physics

The conventional attribution of the entropy cost to the alleged need to erase the Demon's memory arises from the persistent attribution of unphysical classical idealizations to gas molecules, as mentioned in the previous section. The crucially relevant calculation of the entropy cost of measurement, presented in Section 3, can only be seen once these physically inappropriate classical idealizations of gas molecules are dropped, and their quantum nature taken into account.[6] One can identify two aspects of this conventional classical idealization that require elucidation and critical re-evaluation. Both involve violation of the Heisenberg Uncertainty Relation. These are:

(A) The assumption that gas molecules occupy well-defined, phase space point microstates at all times and that the uncertainty about the system's microstate is always epistemic (i.e.,, merely the ignorance of a macro-level observer concerning presumed actually occupied phase space points).

(B) The assumption that a real gas molecule can be localized to a smaller volume (e.g. via insertion of a partition in the containing volume) without entropy cost.

Assumption (A) is the key premise underlying the very possibility of a Maxwell's Demon. In other words, the Demon is the alleged micro-level observer who is presumed to be immune to the macro-level ignorance of the gas' putatively occupied phase space points, and who need only engage in presumed non-entropy increasing actions to ascertain those microstates that are inaccessible to the macro-observer and commence sorting to create a thermodynamic gradient. Thus the Demon's supposed ability to manipulate the gas microstates so as to violate the second law trades precisely on a classical idealization of gas molecules. This assumption is inadvertently illustrated in the ubiquitous, but misleading, representations of gas molecules as little localized particles floating around in their enclosing volumes (as in Figure 1).

On the contrary, and as reminded in the previous section, real gas molecules in confining containers are much more closely approximated by quantum bound states: essentially waves of definite energy and completely indefinite position. For a single molecule, even if we allow for interactions with the box walls, which can introduce decoherence effects, there is no physical justification for modeling the state of the gas molecule as a phase space point, since that violates the uncertainty principle.

---

[6] It should be noted that Norton (2016) provides a quantum version of his Liouville-based proof that no Demon can violate the second law.

Assumption (B), that insertion of a partition does not constitute measurement and costs zero entropy, is traditionally applied to the step in Szilard's engine in which a partition capable of acting as a piston is placed in a chamber, halving the volume occupied by a single gas molecule (but it is assumed that a macro-level observer does not yet know on which side of the chamber it is). Thus the conventional assumption is that the gas molecule is already determinately on or the other side of the box over the time interval during which the partition is placed (i.e., describable by a phase space point), so that there is no disturbance of the molecule during the placement. This unphysical idealization is exemplified in a write- up by K. Yavilberg (2025) which says "when the [partition] is inserted the measurement is not performed yet," despite the fact that his account purportedly addresses a quantum version of the engine. But on the contrary, the partition insertion certainly does function as a measurement in that it forces a localization of the system with respect to the position observable, reducing $\sigma_x$ to ~V/2. By the uncertainty principle, the particle's conjugate momentum spread $\sigma_p$ increases, which from (1) and (2) means that the particle's entropy increases, as already shown in Section 3.

4.2 Erasure of memory devices fails to save second law

Not only has the conventional approach misidentified the source of the entropy increase thwarting the Demon, but the purported source of the entropy increase--erasure of memory degrees of freedom--actually fails to do the job. Norton (e.g. 2005,2011, 2013, 2016, 2017, 2025) and Earman and Norton (e.g., 1999) have already argued that pointing to the alleged need for erasure is not sufficient to save the second law from a Demon, and provided counterexamples (reviewed in Norton, 2011). In these counterexamples, a Demon can carry out a reset of his memory without incurring a significant entropy cost. These counterexamples have not been successfully refuted. Instead they have been merely deflected by reference to computational considerations that are arguably not applicable (Norton 2011, 150). Norton has also provided extensive and detailed analyses (e.g. Norton 2017) showing that thermal fluctuations will thwart devices such as Bennett's "keel and key" construction, proposed as an ostensible way to ascertain the position of the molecule in the cylinder of Szilard's Engine without entropy cost. More generally, Norton's analysis shows that thermal fluctuations will be sources of entropy generation in any attempt to sustain a "reversible" process intended to manipulate micro-level degrees of freedom to disequilibrium states. In any case, as noted above, such devices, when intended to function as measuring instruments, are presumed to carry out mere copying of presumed outcome states that were arrived at without entropy cost. We have already refuted this assumption in Section 3.

Thus, besides the demonstrated failure of Bennett's claim that measurement is always dissipationless copying, it has been shown that the Bennettian recourse to a memory reset is insufficient to save the Second Law. As noted above in reference to Norton (2011), there is in fact no need for an irreversible expansion for the Demon's purpose of preparing his memory for another measurement. The irreversible expansion step may be sufficient for the Demon's memory reset purpose, but it is *not necessary*. Consider the usual example where a gas molecule could be either in the left or right side of a box, with the reset state being on the right. The Demon can simply apply a piston from left to right after every step in which he gains information. If the gas molecule is on the right, it is unchanged and the Demon is ready for the next cycle. If it is on the left, it is pushed to the right, and the Demon is ready for the next cycle. (This is similar to the 'no-erase' reset procedure of Norton, 2011.) This procedure does not involve compression of the phase space, but instead moving the occupied phase space volume from one part of the total phase space to the other about half the time. One might argue that this process leaves a trace of the gas molecules' outcome state somewhere in the system, such that erasure is not complete, but this does not block the conclusion that the Second Law is not saved via "erasure," for two reasons:

(i) Complete erasure not required by the Demon: All that is required, according to Bennett, is that the Demon be able to continue with his sorting given that he has limited memory storage space. And all that he needs in order to do that is to reset his memory degree of freedom. That is accomplished regardless of whether the erasure is "complete" in the sense that no trace remains. There could be a trace left in the system or somewhere else in the Demon's non-information bearing degrees of freedom. That does not stop him from sorting, since his memory degree of freedom remains available; nor does the existence of a trace compensate entropically for the alleged reduction in entropy due to sorting. Thus, pointing to the alleged need for the Demon to reset his memory in order to continue sorting fails to save the second law, since he can do that without sufficient entropy increase to compensate for the entropy decrease due to sorting. Insisting that no trace must remain and that one must therefore employ an irreversible expansion step amounts to an *ad hoc* addition of entropy not actually required for the Demon's stated sorting process (i.e., he could refuse to employ that and still carry out his procedure), and is imposed only to save the second law. Thus, that fails to exorcise the Demon.

(ii) Double standard (full erasure not achieved even in Bennettian erasure): Even if we carry out the erasure by irreversible expansion as in the Bennettian convention, it cannot be claimed that no trace of the initial state remains. This point arises by the very same reasoning Maxwell used in his original Demon scenario: under the conventional assumptions of classicality (i.e., determinate phase space microstates undergoing deterministic evolution), a sufficiently tiny agent could discover those allegedly occupied phase space point-microstates. And under these assumptions, the history of the system could be uncovered by such an agent, along with the supposedly "erased" information, which is recorded in the trajectories of the microstates. Thus, under the assumed classical conditions of the scenario and the alleged ability of a Demon to discover microstates without disturbing them--as is insisted upon in the Bennettian tradition--no microstate information is ever really erased, even under an irreversible expansion. The basic point is that erasing "information" by turning it into "heat" never really happens in the classical picture, since "heat" is taken merely as inaccessible information (where the latter is observer-relative). Thus, ironically, complete erasure is not actually accomplished according to the rules assumed in the "erasure" tradition itself. So it will not do to reject the reset procedure discussed in (i) based on the criticism that erasure is not complete.

The conclusion is that asserting the need for erasure via irreversible expansion is merely a way of helping oneself to a compensating entropy increase that need not really be incurred by the Demon under the Bennettian convention (i.e., that measurement is dissipationless copying and that entropy is equivalent to inaccessible but physically instantiated information). And in any case, under the same set of assumptions (essentially idealized classicality and disregarding unpredictable fluctuations), the irreversible expansion invoked for the entropy increase does not really accomplish the claimed full erasure: a demon could always retrace the history of the microstates without entropy cost, so that the information never really goes away. Thus the whole notion of erasure as applied to the memory of a Maxwell's Demon is spurious.

### 4.3 Experimental corroboration of entropy cost of measurement misconstrued as a cost of erasure

An instance of the dysfunctional state of the literature on Maxwell's Demon is a case in which an experiment clearly demonstrates an entropy cost of measurement in the operation of a quantum-level "demon" but is mischaracterized as corroborating a purported entropy cost of erasure. This instance is found in a *Physics Today* article by Lutz and Ciliberto (2015). The authors recount the paradox of Maxwell's Demon and then say: "The proper resolution of the paradox wouldn't come for another 115 years," where their purported "proper resolution" is the conventional account invoking memory erasure. However, the authors then cite an experiment by Raizen (Raizen, 2009) in which the entropy cost comes from the measurement step. The experiment involves atoms and an optical beam constituting a quantum example of the "Demon"; the experimenters and Lutz and Ciliberto correctly note that the arrangement does not violate the Second Law. But the physics of the experiment and even the authors' own discussion of it clearly demonstrate that *it is the measurement process associated with information gain*, *not erasure*, that saves the Second Law. Their description of the experiment is as follows:

"In the [experiment of Raizen, 2009 ], the optical potential serves as the demon. If an atom is determined to be moving from right to left—that is, if it encounters the excitation beam first, the trapdoor is opened. For all other atoms, the trapdoor is closed. Information about the position and internal state of the atoms is stored in the photons scattered by the atoms. **Each time an atom scatters a photon, the entropy of the optical beam increases, because a photon that was propagating coherently with the beam gets scattered in an uncertain direction**."

The bolded sentence (my emphasis) describes the *measurement* of an atom, not an erasure of stored information. Thus, though the authors talk about "storage" of "information" (in conformance with the information-theoretic tradition), the scattering of the beam that they acknowledge as the source of the entropy cost is t*he measurement of the atom's state*, not an erasure of any "stored information. What is particularly striking is that the authors' stated intent is to argue that the need for erasure is the "proper resolution" of the paradox. Instead it is clear that in the referenced experiment, the Demon is foiled not by any erasure step but by the entropy cost associated with measurement of the atom.

5. Conclusion

We have pointed out that real gas molecules are quantum systems, and shown that the measurements needed to make use of them in such devices as the Szilard engine are the actual, physical source of the entropy cost associated with creating the disequilibrium state intended to violate the second law. This fact is a consequence of the entropic uncertainty relation (1) and the understanding that entropy corresponds to the spread of the momentum wavefunction as in (2). This result vindicates the initial view of Brillouin (1951) that it is the measurement step that incurs an entropy cost, although (as noted by Norton, 2025) Brillouin's attempt to cast the physical processes into information-theoretic terms arguably diluted his point and opened the door for counterproductive digressions into information-theoretic accounts that obscured, rather than illuminated, the relevant physics.

Furthermore, despite the repeated claims that the "erasure" account is empirically corroborated, it is actually undermined in specific experiments such as that of Raizen (2009). This situation is a result of the longstanding misidentification of measurement with mere copying, which results from the inappropriate classical idealizations (A) and (B) discussed in Section 4. The conventional view, as repeated by Lutz and Ciliberto (2015) is that "Bennett showed…that there is no fundamental energetic limitation on the measurement process." However, Bennett did nothing of the sort, since he erroneously modeled measurement as mere copying--a classical process. On the contrary, we have shown by way of the quantum-level analysis of Section 3 that "gaining information" about real physical systems like gas molecules always involves an entropy cost. Denial of this fact is based on mischaracterizing real physical systems as idealized classical particles and neglecting the uncertainty principle.

Thus, the tradition of exorcising Maxwell's Demon by appealing to an alleged need for memory erasure, and an associated entropy cost, is fatally flawed. The actual source of entropy that foils the demon is the measurement process. We thus concur with Norton's overall assessment (e.g., Norton 2025) but offer a quantum-level analysis of the precise source of entropy increase in terms of the dissipation involved in ascertaining values of observables. Norton's criticism focuses on the failure of the "erasure" program to take into account fluctuations, and this is a valid point. It is certainly the case, as Norton says, that:

> "What has been overlooked, repeatedly, is that thermal fluctuations preclude the completion of any molecular-scale process, whether it implements a logically reversible computation or anything else. These fluctuations disrupt their completion unless we employ entropically costly procedures to suppress fluctuations." (Norton 2025, 5-6)

What we have done here is to trace those fluctuations to their fundamental source at the quantum level; a molecular-scale process such as manipulating a molecule into one or the other side of a chamber is inherently dissipative in view of the uncertainty principle. We "suppress" its freedom to be on or the other side of the chamber only by an entropy expenditure corresponding to its increase in momentum uncertainty. Specifically, the Entropic Uncertainty Relation (1) prevents any ascertainment of the position of a molecule sufficient to produce work (as in Szilard's engine) without a compensating entropy increase of $\Delta S = -k\ ln\ a$ where $a$ is the factor by which the molecule's position uncertainty is reduced. Thus, if we place a partitioning piston in the center of the chamber of Szilard's engine in order to confine the molecule to one side or the other and thus obtain work from it as it pushes the piston back out, we have already expended $\Delta S = k \ln 2$ before we even find out which side it is on. As is pointed out in Kastner and Schlatter (2024) in a quantitative calculation, the original speed-measuring demon is foiled by the uncertainty principle in that measurement of a molecule's speed leaves its position so uncertain as to prevent the demon from getting it through the door. The demon must measure in order to sort or to extract work, and he is always foiled by the uncertainty principle, either by the entropy cost of extracting the work in the case of the "pressure-demon," or by being unable to sort in the case of the "speed-demon."

Thus, the Bennettian tradition of modeling "information acquisition" as mere copying of already-determinate states of microscopic systems that were arrived at with no entropy increase is untenable, in view of the fact that real physical systems are quantum systems. The second law is saved from Maxwell's Demon by quantum mechanics, not by computational considerations that demonstably (by the present analysis) have little or nothing to do with the relevant physics. In Norton's words:

"The whole episode [of casting the Maxwell's Demon question into information-theoretic terms] indicates a shift of foundational commitments, driven by little more than an overreaching attempt to promote computational conceptions. There are no novel experiments driving the change. We are to suppose that the giants—von Neumann!—were just confused or negligent and that a little more thought completely overturns their insights. My note here reviews how precisely the same problem afflicts Bennett and Landauer's analysis. This is not a literature with stable foundations." (Norton 2025, 4)

In this regard, recall the so-called "von Neumann entropy", which is really a form of information: $I_{VN} = -tr\ \rho ln\rho,$ where ρ is the density operator for a system. While it has been argued here that we should not indentify "information" in general with thermodynamic entropy, they can be related, as discussed in Kastner and Schlatter (2024) and through (1); indeed a quantity of von Neumann information is what appears in each term of (1). It is well known that in general, the von Neumann Information increases in a measurement (e.g. Gaspard 2013), and this is related to the uncertainty principle. When position is measured, the conjugate von Neumann information associated with momentum corresponds to entropy as in (2); thus its well-known increase constitutes the increase of a system's entropy upon a position measurement. The Bennettian tradition of discounting the effects of measurement by a Demon completely disregards this crucial point without ever having refuted it.

It is the present author's hope to have provided, along with Norton's rightful emphasis on the literature's fatal neglect of thermal fluctuations, a bit more of a firm foundation for the analysis of Maxwell's demon and related issues.